\documentclass[a4,11pt,reqno]{amsart}%
\usepackage[english]{babel}
\usepackage[latin1]{inputenc}
\usepackage{graphics}
\usepackage{ulem}
\usepackage{hhline}
\usepackage{dsfont}
\usepackage{mathrsfs}
\usepackage{fancyhdr}
\usepackage{amsmath,amssymb}
\usepackage{rotating}
\usepackage[latin1]{inputenc}
\usepackage[T1]{fontenc}
\usepackage{fancybox}
\usepackage{color}
\usepackage{colortbl}
\usepackage{setspace}
\usepackage{enumerate}
\usepackage[nice]{nicefrac}
\usepackage{amsthm}
\usepackage{multicol}
\usepackage{pifont}
\usepackage[french]{minitoc}
\usepackage{latexsym,amsfonts}
\usepackage{amsthm}
\usepackage{varioref}
\usepackage{textcomp}
\usepackage{lmodern}
\usepackage{mathpazo}
\usepackage{euscript}
\usepackage[pdftex]{hyperref}
\numberwithin{equation}{section}
\makeatletter\@addtoreset{equation}{section}

\DeclareMathSymbol{\subsetneqq}{\mathbin}{AMSb}{36}

\usepackage[pdftex]{hyperref}
\newcommand{\fin}{\hfill {$\Box$}}

\newtheorem {theorem}{Theorem}[section]

\newtheorem {definition}[theorem]{Definition}

\newtheorem {proposition}[theorem]{Proposition}

\numberwithin{equation}{section}
\makeatletter\@addtoreset{equation}{section}

\begin{document}

\title[Class of coherent states with Meixner-Pollaczek polynomials]{A new class of coherent states
with Meixner-Pollaczek polynomials for the Gol'dman-Krivchenkov Hamiltonian}
\author[Z. Mouayn]{Zouha\"{i}r Mouayn}
\address{Department of Mathematics, Faculty of Sciences and Technics (M'Ghila), PO.
Box 523,
Sultan My Slimane University, CP. 23 000, B\'{e}ni Mellal, Morocco}
\email{mouayn@fstbm.ac.ma}
\begin{abstract}
A class of generalized coherent states with a new type of the identity
resolution are constructed by replacing the labeling parameter $z^{n}/\sqrt{%
n!}$ of the canonical coherent states by Meixner-Pollaczek polynomials with
specific parameters. The constructed coherent states belong to the state
Hilbert space of the Gol'dman-Krivchenkov Hamiltonian.
\end{abstract}

\maketitle

\section{Introduction}

Coherent states are mathematical tools which provide a close connection
between classical and quantum formalisms and have essentially many
definitions. In general, coherent states are a specific overcomplete family
of vectors in the Hilbert space of the problem that describes the quantum
phenomena and solves the identity of this Hilbert space \cite{1}.

The canonical coherent states for the harmonic oscillator have long been
known and their properties have frequently been
taken as models for defining the notion of coherent states for other models
\cite{2,5}.

In this paper, we are concerned with the model of the Gol'dman-Krivchenkov
Hamiltonian \cite{6} acting on the Hilbert space of square integrable
functions on the positive real half-line. We, precisely construct a family
of coherent states labeled by points of the whole real line, depending on
some parameters and belonging to the state Hilbert space of this
Hamiltonian. To achieve this, we will adopt a formalism of canonical
coherent states when written as superpositions of the harmonic oscillator
number states. That is, we present our generalized coherent states as a
superposition of an orthonormal basis of the state Hilbert space of the
Gol'dman-Krivchenkov Hamiltonian, whose identity is solved by a new way. We
choose the Meixner-Pollaczek polynomials to play the role of coefficients in
this superposition. This choice enables us to present the constructed states
in a closed form.

The paper is organized as follows. In Section 2, we recall briefly some
needed spectral properties of the Gol'dman-Krivchenkov Hamiltonian. Section
3 is devoted to the coherent states formalism we will be using. This
formalism is applied in Section 4 so as to construct a family of coherent
states in the state Hilbert space of the Hamiltonian we are dealing with. In
Section 5 we conclude with a summary.

\section{The Gol'dman-Krivchenkov Hamiltonian}

An anharmonic potential that can be used to calculate the vibrational
energies of a diatomic molecules has the form
\begin{equation}
V_{\varrho ,\kappa _{0}}(\xi ):=\varrho \left( \frac{\xi }{\kappa _{0}}-%
\frac{\xi _{0}}{\xi }\right) ^{2}  \label{2.1}
\end{equation}
where $\kappa _{0}$ $>0$ denotes the equilibrium bond length which is the
distance between the diatomic nuclei, and \ $\varrho >0$ with $F=\varrho
\kappa _{0}^{-2}$ represents a constant force. The associated stationary Schr%
ödinger equation reads
\begin{equation}
-\frac{d^{2}}{d\xi ^{2}}\psi (\xi )+\varrho \left( \frac{\xi }{\kappa _{0}}-%
\frac{\kappa _{0}}{\xi }\right) ^{2}\psi (\xi )=\lambda \psi (\xi ),
\label{2.2}
\end{equation}
with $\psi (0)=0,$ namely $\psi $ satisfies the Dirichlet boundary
condition. It is an exactly solvable equation. Indeed, according to
\cite[p. 11288]{7}, the energy spectrum is given by
\begin{equation}
\lambda _{m}^{\varrho ,\kappa _{0}}:=4\kappa _{0}^{-1}\sqrt{\varrho }\left(
m+\frac{1}{2}+\frac{1}{4}\left( \sqrt{1+4\varrho \kappa _{0}^{2}}-2\kappa
_{0}\sqrt{\varrho }\right) \right) , \quad m=0,1,2, \cdots   \label{2.3}
\end{equation}
whereas the wave functions of the exact solutions of Eq. \eqref{2.2} take
the form
\begin{equation}
<\xi \mid \psi _{m}^{\varrho ,\kappa _{0}}>\propto \xi ^{q}\exp \left( -%
\frac{\sqrt{\varrho }}{2\kappa _{0}}\xi ^{2}\right) \text{ }_{1}\digamma
_{1}\left( -m,q+1,\frac{\sqrt{\varrho }}{2\kappa _{0}}\xi ^{2}\right) ,
\label{2.4}
\end{equation}
where $q=\frac{1}{2}\left( 1+\sqrt{1+4\varrho \kappa _{0}^{2}}\right) $ and $%
_{1}\digamma _{1}\left( .\right) $ denotes the confluente hypergeometric
function which can also be expressed in terms of Laguerre polynomials as
\cite[p. 240]{8}:
\begin{equation}
_{1}\digamma _{1}\left( -m,q+1,u\right) =\frac{m!}{(q)_{m}}L_{m}^{(q)}(u)
\label{2.5}
\end{equation}
in terms of the Pochhammer symbols
\begin{equation}
(a)_{0}=1,(a)_{m}=a(a+1)  \cdots (a+m-1) =\frac{\Gamma
(a+m) }{\Gamma (a)}, \quad m=1,2, \cdots .  \label{2.6}
\end{equation}
To simplify the notation, we introduce the new parameters $\alpha :=\varrho
\kappa _{0}^{2}$ and $\beta :=\kappa _{0}^{-1}\sqrt{\varrho }$, and thereby
the Hamiltonian in Eq. \eqref{2.2} takes the form
\begin{equation}
\Delta _{\alpha ,\beta }:=-\frac{d^{2}}{d\xi ^{2}}+\beta ^{2}\xi ^{2}+\frac{%
\alpha }{\xi ^{2}}, \quad \xi \in \mathbb{R}_{+},\beta ,\alpha >0  \label{2.7}
\end{equation}
called Gol'dman-Krivchenvov Hamiltonian (\cite[p. 11288]{7}). Its spectrum in\ the Hilbert space $L^{2}\left( \mathbb{R}%
_{+},d\xi \right) $\ reduces to a discrete part consisting of eigenvalues of
the form  (\cite[pp. 9-10]{9}):
\begin{equation}
\lambda _{m}^{\gamma ,\beta }:=2\beta ( 2m+\gamma (\alpha)) ,\gamma =\gamma (\alpha) =1+\frac{1}{2}\sqrt{1+4\alpha }%
\quad , m=0,1,2, \cdots ,  \label{2.8}
\end{equation}
and wavefunctions of the corresponding normalized eigenfunctions are given
by
\begin{equation}
<\xi \mid \psi _{m}^{\gamma ,\beta }>:=\left( \frac{2\beta ^{\gamma }m!}{%
\Gamma \left( \gamma +m\right) }\right) ^{\frac{1}{2}}\xi ^{\gamma -\frac{1}{%
2}}e^{-\frac{1}{2}\beta \xi ^{2}}L_{m}^{\left( \gamma -1\right) }\left(
\beta \xi ^{2}\right) , \quad m=0,1,2, \cdots  \label{2.9}
\end{equation}
The set of functions in (2.9)  constitutes a complete
orthonormal basis for the Hilbert space $L^{2}(\mathbb{R}_{+},d\xi ).$

\textbf{Remark 2.1. }We should note that the eigenvalues in (2.8) together with their eigenfunctions could also be
obtained by using raising and lowering operators throughout a factorization
of the Hamiltonian $\Delta _{\alpha ,\beta }$ in $\left( 2.7\right) $ based
on the Lie algebra $su(1,1)$ commutation relations  \cite[pp. 3-4]{10}.

\section{A coherent states formalism}

In this section, we adopt a new generalization of the canonical coherent
states, which extend a well known generalization (\cite[p. 4568]{11}) by
considering a kind of the identity resolution that we obtain as a limit with
respect to a certain parameter. Precisely, we propose the following
definition:

\begin{definition}
Let $\mathcal{H}$ be a separable Hilbert space with an orthonormal basis $%
\left\{ \psi _{n}\right\} _{n=0}^{+\infty }.$ Let $\frak{D}$ $\subseteq %
\mathbb C$ be an open subset of $\mathbb C$ and let $\Phi _{n}:\frak{%
D\rightarrow }\mathbb C,n=0,1,2,\cdots $ be a sequence of complex functions.
Define
\begin{equation}
\mid z,\varepsilon >:=\left( N_{\varepsilon }(z)\right) ^{-\frac{1}{2}%
}\sum\limits_{n=0}^{+\infty }\frac{\Phi _{n}(z)}{\sqrt{\sigma _{\varepsilon
}(n)}}\mid \psi _{n}>, \quad z\in \frak{D}, \ \ \varepsilon >0,  \label{3.1}
\end{equation}
where $N_{\varepsilon }(z)$ is a normalization factor and $\sigma
_{\varepsilon }(n),n=0,1,2,\cdots $ a sequence of positive numbers depending
on $\varepsilon >0$. The set of vectors $\left\{ \mid z,\varepsilon >,z\in
\frak{D}\right\} $ is said to form a set of generalized coherent states if :%
\newline
$\left( i\right) $ for each fixed $\varepsilon >0$ and $z\in \frak{D,}$ the
state $\mid z,\varepsilon >$ is normalized, that is $<z,\varepsilon \mid
z,\varepsilon >_{\mathcal{H}}=1,$\newline
$\left( ii\right) $ the states $\left\{ \mid z,\varepsilon >,z\in \frak{D}%
\right\} $ satisfy the following resolution of the identity
\begin{equation}
\lim_{\varepsilon \rightarrow 0^{+}}\int\limits_{\frak{D}}\mid z,\varepsilon
><z,\varepsilon \mid d\mu _{\varepsilon }(z)=\mathbf{1}_{\mathcal{H}}
\label{3.2}
\end{equation}
where $d\mu _{\varepsilon }$ is an appropriately chosen measure and $\mathbf{%
1}_{\mathcal{H}}$ is the identity operator on the Hilbert space $\mathcal{H}.
$
\end{definition}

We should precise that, in the above definition, the Dirac's\textit{\ bra-ket%
} notation $\mid z,\varepsilon ><z,\varepsilon \mid $ means the
rank-one-operator $\varphi \longmapsto <\varphi \mid z,\varepsilon >_{%
\mathcal{H}}\mid z,\varepsilon >,$ $\varphi \in \mathcal{H}.$ Also, the
limit in $\left( ii\right) $ is to be understood as follows. Define the
operator
\begin{equation}
\mathcal{O}_{\varepsilon }[\varphi] \left( \cdot \right)
:=\left( \int\limits_{\frak{D}}\mid z,\varepsilon ><z,\varepsilon \mid d\mu
_{\varepsilon }(z)\right) [\varphi] \left( \cdot \right)
\label{3.3}
\end{equation}
then the above limit (3.2) means that $\mathcal{O}_{\varepsilon }\left[
\varphi \right] \left( \cdot \right) \rightarrow $ $\varphi \left( \cdot
\right) $ as $\varepsilon \rightarrow 0^{+},$ \textit{almost every where }%
with respect to $\left( \cdot \right) .$

\textbf{Remark 3.1}. The formula $\left( 3.1\right) $\ can be considered as
a generalization of the series expansion of the canonical coherent states
\begin{equation}
\mid z>:=\left( e^{\left| z\right| ^{2}}\right) ^{-\frac{1}{2}%
}\sum_{n=0}^{+\infty }\frac{z^{k}}{\sqrt{n!}}\mid \phi _{n}>,z\in \mathbb C
\label{3.4}
\end{equation}
with $\phi _{n}$, $n=0,1,2, \cdots $\ being an orthonormal basis in $L^{2}\left( %
\mathbb{R},d\xi \right) $ of eigenstates of the harmonic oscillator, which
is given by the functions $\phi _{n}(\xi ):=\left( \sqrt{\pi }2^{n}n!\right)
^{-\frac{1}{2}}e^{-\frac{1}{2}\xi ^{2}}H_{n}(\xi )$ where $H_{n}(\cdot) $ denotes $n$th Hermite polynomial (\cite[p. 249]{8}).

\section{Coherent states attached to $\Delta _{\alpha ,\beta }$}

As announced in section 1, we now will construct a set of the normalized
states labeled by points $x\in \mathbb{R}$ and depending on the parameters: $%
\theta \in \left] 0,\pi \right[ ,\gamma >1$, $\beta >0$ and $\varepsilon >0.
$ These states will be denoted $\mid x,\varepsilon >_{\theta ,\gamma ,\beta
} $ and will belong to $L^{2}(\mathbb{R}_{+},d\xi) $ the state
Hilbert space of the Hamiltonian $\Delta _{\alpha ,\beta }$ in (2.7).

\begin{definition}
Define a set of states $\left( \mid x,\varepsilon >_{\theta ,\gamma ,\beta
}\right) _{x\in \mathbb{R}}$ labeled by points $x\in \mathbb{R}$ and
depending on the parameters $\theta \in \left] 0,\pi \right[ ,\gamma >1,$ $%
\beta >0$ and $\varepsilon >0$ by
\begin{equation}
\mid x,\varepsilon >_{\theta ,\gamma ,\beta }:=\left( \mathcal{N}_{\theta
,\gamma ,\beta ,\varepsilon }(x)\right) ^{-\frac{1}{2}}\sum\limits_{m=0}^{+%
\infty }\frac{P_{m}^{\left( \frac{1}{2}\gamma \right) }\left( x,\theta
\right) }{\sqrt{\sigma _{\varepsilon }^{\beta ,\gamma }(m) }}%
\mid \psi _{m}^{\gamma ,\beta }>  \label{4.1}
\end{equation}
with the precisions:\newline
$\smallskip \bullet \mathcal{N}_{\theta ,\gamma ,\beta ,\varepsilon }\left(
x\right) $ is a normalization factor such that $_{\theta ,\gamma ,\beta }$ $%
<x,\varepsilon \mid x,\varepsilon >_{\theta ,\gamma ,\beta }=1$\newline
$\bullet P_{m}^{\left( \frac{1}{2}\gamma \right) }(x,\theta) $
are the Meixner-Pollaczek polynomials given by\newline
\begin{equation}
P_{m}^{\left( \frac{1}{2}\gamma \right) }(x,\theta) =\frac{%
(\gamma )_{m}}{m!}e^{im\theta }{}_{2}F_{1}(-m,\frac{\gamma }{2}+ix,\gamma
,1-e^{-2i\theta })  \label{4.2}
\end{equation}
where $_{2}F_{1}(.)$ denotes the Gauss hypergeometric function.\newline
$\bullet \sigma _{\varepsilon }^{\beta ,\gamma }(m) ,$ $%
m=0,1,2,\cdots, $ are sequences of positive numbers given by
\begin{equation}
\sigma _{\varepsilon }^{\beta ,\gamma }(m) :=\left( m!\right)
^{-1}(\gamma )_{m}e^{2\beta (2m+\gamma) \varepsilon },
\label{4.3}
\end{equation}
with $\gamma =\gamma (\alpha) =1+\frac{1}{2}\sqrt{1+4\alpha }$%
\newline
$\bullet \mid \psi _{m}^{\gamma ,\beta }>,m=0,1,2,\cdots $ , is the
orthonormal basis of $L^{2}(\mathbb{R}_{+},d\xi) $ given in $%
\left( 2.6\right) $
\end{definition}

We shall give the main properties on these states in the following three
propositions.

\begin{proposition}
Let $\theta \in \left] 0,\pi \right[ ,\gamma >1$ and $\varepsilon >0$ be
fixed parameters. Then, the normalization factor in (4.1) has the following
expression:
\begin{equation}
\mathcal{N}_{\theta ,\gamma ,\beta ,\varepsilon }(x)=\frac{\left(
1-e^{-4\varepsilon \beta +2i\theta }\right) ^{2ix}}{\left( 2sh2\varepsilon
\beta \right) ^{\gamma }}\times _{2}F_{1}\left( \frac{1}{2}\gamma +ix,\frac{1%
}{2}\gamma +ix,\gamma ;\frac{-4e^{-4\varepsilon \beta }\sin ^{2}\theta }{%
\left( 1-e^{-4\varepsilon \beta }\right) ^{2}}\right)   \label{4.4}
\end{equation}
for every $x\in \mathbb{R}$.
\end{proposition}

\noindent {\it Proof.}
calculate the normalization factor, we start by writing the condition
\begin{equation}
1=_{\theta ,\gamma ,\beta }<x,\varepsilon \mid x,\varepsilon >_{\theta
,\gamma ,\beta }  \label{4.5}
\end{equation}
Eq.(4.5) is equivalent to
\begin{equation}
\left( \mathcal{N}_{\theta ,\gamma ,\beta ,\varepsilon }(x)
\right) ^{-1}\sum\limits_{m=0}^{+\infty }\frac{1}{\sigma _{\varepsilon
}^{\beta ,\gamma }(m) }\left( P_{m}^{\left( \frac{1}{2}\gamma
\right) }(x,\theta) \right) ^{2}=1,  \label{4.6}
\end{equation}
Inserting the expression $\left( 4.3\right) $ into Eq.$\left( 4.6\right) ,$
we obtain that
\begin{equation}
\mathcal{N}_{\theta ,\gamma ,\varepsilon }(x) =e^{-2\varepsilon
\beta \gamma }\sum\limits_{m=0}^{+\infty }\frac{m!}{(\gamma)
_{m}}\left( e^{-4\varepsilon \beta }\right) ^{m}\left( P_{m}^{\left( \frac{1%
}{2}\gamma \right) }(x,\theta) \right) ^{2}  \label{4.7}
\end{equation}
Next, we make use of the following identity (\cite[p. 527]{12}):
\begin{align}
\sum\limits_{m=0}^{+\infty }&\frac{m!}{(\gamma) _{m}}\mu
^{m}P_{m}^{\left( \frac{1}{2}\gamma \right) }\left( x,\theta _{1}\right)
P_{m}^{\left( \frac{1}{2}\gamma \right) }\left( y,\theta _{2}\right) =\left(
1-\mu e^{i\left( \theta _{1}-\theta _{2}\right) }\right) ^{-\frac{1}{2}%
\gamma -iy}  \left( 1-\mu e^{i\left( \theta _{2}-\theta _{1}\right) }\right) ^{-%
\frac{1}{2}\gamma -ix}
\label{4.8} \\ &
\times \left( 1-\mu e^{i\left( \theta _{1}+\theta _{2}\right)
}\right) ^{ix+iy}  {_{2}F_{1}}\left( \frac{1}{2}\gamma +ix,\frac{1}{2}\gamma +ix,\gamma ;%
\frac{-4\mu \sin \theta _{1}\sin \theta _{2}}{\left( 1-\mu e^{i\left( \theta
_{2}-\theta _{1}\right) }\right) \left( 1-\mu e^{i\left( \theta _{1}-\theta
_{2}\right) }\right) }\right)
\nonumber \end{align}
for $\theta _{1}=\theta _{2}=\theta $, $x=y$ and $\mu =e^{-4\varepsilon
\beta }$. Then, we arrive at the result
\begin{equation}
\mathcal{N}_{\theta ,\gamma ,\beta ,\varepsilon }(x) =\frac{%
\left( 1-e^{-4\varepsilon \beta +2i\theta }\right) ^{2ix}}{\left(
2sh2\varepsilon \beta \right) ^{\gamma }}\times _{2}F_{1}\left( \frac{1}{2}%
\gamma +ix,\frac{1}{2}\gamma +ix,\gamma ;\frac{-4e^{-4\varepsilon \beta
}\left( \sin \theta \right) ^{2}}{\left( 1-e^{-4\varepsilon \beta }\right)
^{2}}\right)   \label{4.9}
\end{equation}
This ends the proof.
\fin \\

Now, we will present a closed form for the constructed generalized coherent
states as follows:

\begin{proposition}
Let $\theta \in \left] 0,\pi \right[ ,\gamma >1$, $\beta >0$ and $%
\varepsilon >0$ be fixed parameters. Then, the wave functions of the states $%
\mid x,\varepsilon >_{\theta ,\gamma ,\beta }$ defined in \textit{(4.1) }can
be written in a closed form as
\begin{align}
<\xi \mid x,\varepsilon >_{\theta ,\gamma ,\beta }&=\frac{\sqrt{2\beta
^{\gamma }}e^{-\varepsilon \beta \gamma }}{\sqrt{\Gamma \left( \gamma
\right) }}\left| 1-e^{-2\beta \varepsilon +i\theta }\right| ^{-\gamma
}\left( \frac{1-e^{-2\beta \varepsilon +i\theta }}{1-e^{-2\beta \varepsilon
-i\theta }}\right) ^{ix}\left( \mathcal{N}_{\theta ,\gamma ,\beta
,\varepsilon }(x)\right) ^{-\frac{1}{2}}  \label{4.10} \\ &
\times \xi ^{\gamma -\frac{1}{2}}\exp \left( -\frac{1}{2}\beta \xi
^{2}\left( \frac{1+e^{-2\beta \varepsilon +i\theta }}{1-e^{-2\beta
\varepsilon +i\theta }}\right) \right) \times _{1}\digamma _{1}\left( \frac{%
\gamma }{2}+ix,\gamma ;\frac{2i\beta \xi ^{2}\sin \theta e^{-2\beta
\varepsilon }}{\left| 1-e^{-2\beta \varepsilon +i\theta }\right| ^{2}}%
\right)   \nonumber
\end{align}
for every $\xi \in \mathbb{R}_{+}.$
\end{proposition}

\noindent {\it Proof.}
start by writing the expression of the wave function of states $\mid
x,\varepsilon >_{\theta ,\gamma ,\beta }$ according to definition $(4.1)$ as
\begin{equation}
<\xi \mid x,\varepsilon >_{\theta ,\gamma ,\beta }=\left( \mathcal{N}%
_{\theta ,\gamma ,\beta ,\varepsilon }(x) \right) ^{-\frac{1}{2}%
}\sum\limits_{m=0}^{+\infty }\frac{P_{m}^{\left( \frac{1}{2}\gamma \right)
}(x,\theta) }{\sqrt{\sigma _{\varepsilon }^{\theta ,\gamma
}(m) }}<\xi \mid \psi _{m}^{\gamma ,\beta }>,\xi \in \mathbb{R}%
_{+}.  \label{4.11}
\end{equation}
We have thus to look for a closed form of the series
\begin{equation}
\mathcal{S}(\xi) :=\sum\limits_{m=0}^{+\infty }\frac{%
P_{m}^{\left( \frac{1}{2}\gamma \right) }(x,\theta) }{\sqrt{%
\sigma _{\varepsilon }^{\beta ,\gamma }(m) }}<\xi \mid \psi
_{m}^{\gamma ,\beta }>  \label{4.12}
\end{equation}
which also reads
\begin{align}
\mathcal{S}(\xi) &=\sum\limits_{m=0}^{+\infty }\frac{\sqrt{m!}}{%
\sqrt{(\gamma) _{m}}}e^{-\varepsilon \beta \left( 2m+\gamma
(\alpha) \right) }P_{m}^{\left( \frac{1}{2}\gamma \right)
}(x,\theta) <\xi \mid \psi _{m}^{\gamma ,\beta }>  \label{4.13}
 \\ &
=e^{-\varepsilon \beta \gamma (\alpha)
}\sum\limits_{m=0}^{+\infty }\frac{\sqrt{m!}}{\sqrt{(\gamma)
_{m}}}e^{-m2\beta \varepsilon }P_{m}^{\left( \frac{1}{2}\gamma \right)
}(x,\theta) <\xi \mid \psi _{m}^{\gamma ,\beta }>.  \label{4.14}
\end{align}
Replacing the Meixner-Pollaczeck polynomial $P_{m}^{\left( \frac{1}{2}\gamma
\right) }(x,\theta) $ by its expression in terms of the Gauss
hypergeometric function as given in $\left( 4.2\right) $, then Eq.$\left(
4.14\right) $ takes the form
\begin{equation}
\mathcal{S}(\xi) =e^{-\varepsilon \beta \gamma \left( \alpha
\right) }\sum\limits_{m=0}^{+\infty }\frac{\sqrt{(\gamma) _{m}}%
}{\sqrt{m!}}\left( e^{-2\beta \varepsilon +i\theta }\right) ^{m}\text{ }%
_{2}F_{1}(-m,\frac{\gamma }{2}+ix,\gamma ,1-e^{-2i\theta })<\xi \mid \psi
_{m}^{\gamma ,\beta }>  \label{4.15}
\end{equation}
Put $\tau :=e^{-2\beta \varepsilon +i\theta },\left| \tau \right|
=e^{-2\beta \varepsilon }<1.$ Then Eq.$\left( 4.15\right) $ becomes
\begin{equation}
\mathcal{S}(\xi) =e^{-\varepsilon \beta \gamma \left( \alpha
\right) }\sum\limits_{m=0}^{+\infty }\frac{\sqrt{(\gamma) _{m}}%
}{\sqrt{m!}}\tau ^{m}\text{ }_{2}F_{1}(-m,\frac{\gamma }{2}+ix,\gamma
,1-e^{-2i\theta })<\xi \mid \psi _{m}^{\gamma ,\beta }>.  \label{4.16}
\end{equation}
Replacing the wavefunction $<\xi \mid \psi _{m}^{\gamma ,\beta }>$ by its
expression in $\left( 2.9\right) $, we get that
\begin{align}
\mathcal{S}(\xi) &=e^{-\varepsilon \beta \gamma \left( \alpha
\right) }\sum\limits_{m=0}^{+\infty }\frac{\sqrt{(\gamma) _{m}}%
}{\sqrt{m!}}\tau ^{m}\text{ }_{2}F_{1}(-m,\frac{\gamma }{2}+ix,\gamma
,1-e^{-2i\theta })
 \label{4.17}\\ &
\times \left( \frac{2\beta ^{\gamma }m!}{\Gamma \left( \gamma +m\right) }%
\right) ^{\frac{1}{2}}\xi ^{\gamma -\frac{1}{2}}\exp \left( -\frac{1}{2}%
\beta \xi ^{2}\right) L_{m}^{\left( \gamma -1\right) }\left( \beta \xi
^{2}\right)   \nonumber
\end{align}
Now, we summarize up the above calculations by writing
\begin{equation}
<\xi \mid x,\varepsilon >_{\theta ,\gamma ,\beta }=\frac{\sqrt{2\beta
^{\gamma }}}{\sqrt{\Gamma (\gamma) }}e^{-\varepsilon \beta
\gamma }\left( \mathcal{N}_{\theta ,\gamma ,\beta ,\varepsilon }\left(
x\right) \right) ^{-\frac{1}{2}}\xi ^{\gamma -\frac{1}{2}}e^{-\frac{1}{2}%
\beta \xi ^{2}}\frak{S}_{\gamma ,\beta }^{\tau ,x,\theta }(\xi)
\label{4.18}
\end{equation}
where
\begin{equation}
\frak{S}_{\gamma ,\beta }^{\tau ,x,\theta }(\xi)
:=\sum\limits_{m=0}^{+\infty }\tau ^{m}\text{ }_{2}\digamma _{1}(-m,\frac{%
\gamma }{2}+ix,\gamma ,1-e^{-2i\theta })L_{m}^{\left( \gamma -1\right)
}\left( \beta \xi ^{2}\right)   \label{4.19}
\end{equation}
Next, with the help of the generating formula \cite[p. 213]{13}:
\begin{align}
\sum\limits_{n=0}^{+\infty }t^{n}\text{ }_{2}\digamma _{1}(-n,c,1+\nu
;y)L_{n}^{(\nu) }(u) & =\left( 1-t\right) ^{-1+c-\nu
}\left( 1-t+yt\right) ^{-c}
\label{4.20} \\ &
\times \exp \left( \frac{-ut}{1-t}\right) \text{ }_{1}\digamma _{1}\left(
c,1+\nu ,\frac{yut}{\left( 1-t\right) \left( 1-t+yt\right) }\right) \nonumber
\end{align}
for $t=\tau ,n=m,c=\frac{\gamma }{2}+ix,y=1-e^{-2i\theta },\nu =\gamma -1$
and $u=\beta \xi ^{2},$we obtain an expression of series $\left( 4.19\right)
$ as
\begin{align}
\frak{S}_{\gamma ,\beta }^{\tau ,x,\theta }(\xi) &=\left( 1-\tau
\right) ^{-\frac{\gamma }{2}+ix}\left( 1-e^{-2i\theta }\tau \right)
^{-\left( \frac{\gamma }{2}+ix\right) }\exp \left( -\frac{\beta \zeta
^{2}\tau }{1-\tau }\right)
 \label{4.21} \\ &
\times \exp ( -\frac{\beta \xi ^{2}\tau }{1-\tau })
_{1}\digamma _{1}\left( \frac{\gamma }{2}+ix,\gamma ,\frac{(
1-e^{-2i\theta }) \beta \xi ^{2}\tau }{(1-\tau) (
1-e^{-2i\theta }\tau ) }\right)  \nonumber
\end{align}
Finally, we arrive at the following expression of the wave functions
\begin{align}
<\xi \mid x,\varepsilon >_{\theta ,\gamma ,\beta }&=\frac{\sqrt{2\beta
^{\gamma }}e^{-\varepsilon \beta \gamma }}{\sqrt{\Gamma \left( \gamma
\right) }}\left| 1-e^{-2\beta \varepsilon +i\theta }\right| ^{-\gamma
}\left( \frac{1-e^{-2\beta \varepsilon +i\theta }}{1-e^{-2\beta \varepsilon
-i\theta }}\right) ^{ix}\left( \mathcal{N}_{\theta ,\gamma ,\beta
,\varepsilon }(x) \right) ^{-\frac{1}{2}}
  \label{4.22} \\ &
\times \xi ^{\gamma -\frac{1}{2}}\exp \left( -\frac{1}{2}\beta \xi
^{2}\left( \frac{1+e^{-2\beta \xi +i\theta }}{1-e^{-2\beta \xi +i\theta }}%
\right) \right) \times _{1}\digamma _{1}\left( \frac{\gamma }{2}+ix,\gamma ;%
\frac{2i\beta \xi ^{2}\sin \theta e^{-2\beta \varepsilon }}{\left|
1-e^{-2\beta \varepsilon +i\theta }\right| ^{2}}\right) \nonumber
\end{align}
This ends the proof.
\fin \\

\begin{proposition}
The states $\mid x,\varepsilon >\equiv \mid x,\varepsilon >_{\theta ,\gamma
,\beta },x\in \mathbb{R}$ satisfy the following resolution of the identity
\begin{equation}
\lim_{\varepsilon \rightarrow 0^{+}}\int\limits_{\mathbb{R}}\mid
x,\varepsilon ><\varepsilon ,x\mid d\mu _{\theta ,\gamma ,\varepsilon
}(x) =\mathbf{1}_{L^{2}(\mathbb{R}_{+},d\xi) }
\label{4.23}
\end{equation}
where $\mathbf{1}_{L^{2}(\mathbb{R}_{+},d\xi) }$ is the
identity operator on the Hilbert space $L^{2}\left( \mathbb{R}_{+},d\xi
\right) $ and $d\mu _{\theta ,\gamma ,\varepsilon }(x) $ is a
measure on $\mathbb{R}$ with the expression
\begin{equation}
d\mu _{\theta ,\gamma ,\beta ,\varepsilon }(x) :=\frac{\left(
2\sin \theta \right) ^{\gamma -1}}{\pi ^{\gamma -1}\Gamma \left( \gamma
\right) \cos ec(\theta )}\mathcal{N}_{\theta ,\gamma ,\beta ,\varepsilon
}(x) e^{-\left( \pi -2\theta \right) x}\left| \Gamma \left(
\frac{\gamma }{2}+ix\right) \right| ^{2}dx,  \label{4.24}
\end{equation}
$\mathcal{N}_{\theta ,\gamma ,\beta ,\varepsilon }(x) $ being
the normalization factor given in proposition\textit{\ (4.1)}$.$
\end{proposition}

\noindent {\it Proof.}
Let us assume that the measure takes the form
\begin{equation}
d\mu _{\theta ,\gamma ,\beta ,\varepsilon }(x) =\mathcal{N}%
_{\theta ,\gamma ,\beta ,\varepsilon }(x) \Upsilon _{\theta
,\gamma }(x) dx  \label{4.25}
\end{equation}
where $\Upsilon _{\theta ,\gamma }(x) $ is an auxiliary density
to be determined. Let $\varphi \in L^{2}(\mathbb{R}_{+},d\xi) $
and let us start by writing the following action
\begin{align}
\mathcal{O}_{\theta ,\gamma ,\beta ,\varepsilon }[\varphi]
&:=\left( \int\limits_{\mathbb{R}}\mid x,\varepsilon ><x,\varepsilon \mid
d\mu _{\theta ,\gamma ,\beta ,\varepsilon }(x) \right) \left[
\varphi \right]   \label{4.26}
\\&
=\int\limits_{\mathbb{R}}<\varphi \mid x,\varepsilon ><x,\varepsilon \mid
d\mu _{\theta ,\gamma ,\beta ,\varepsilon }(x)   \label{4.27}
\end{align}
We make use of the definition of $\mid x,\varepsilon >$ given in $\left(
4.1\right) :$
\begin{align}
\mathcal{O}_{\theta ,\gamma ,\beta ,\varepsilon }[\varphi]
&=\int\limits_{\mathbb{R}}\Big<\varphi ,\left( \mathcal{N}_{\theta ,\gamma ,\beta
,\varepsilon }(x) \right) ^{-\frac{1}{2}}\sum\limits_{m=0}^{+%
\infty }\frac{P_{m}^{\left( \frac{1}{2}\gamma \right) }\left( x,\theta
\right) }{\sqrt{\sigma _{\varepsilon }^{\beta ,\gamma }(m) }}%
\Big| \psi _{m}^{\gamma ,\beta }\Big><x,\varepsilon \mid d\mu _{\theta ,\gamma
,\beta ,\varepsilon }(x)   \label{4.28}
\\&
=\int\limits_{\mathbb{R}}\sum\limits_{m=0}^{+\infty }\frac{P_{m}^{\left(
\frac{1}{2}\gamma \right) }(x,\theta) }{\sqrt{\sigma
_{\varepsilon }^{\beta ,\gamma }(m) }}<\varphi \mid \psi
_{m}^{\gamma ,\beta }><x,\varepsilon \mid \left( \mathcal{N}_{\theta ,\gamma
,\beta ,\varepsilon }(x) \right) ^{-\frac{1}{2}}d\mu _{\theta
,\gamma ,\beta ,\varepsilon }(x)   \label{4.29}
\\&
=\left( \sum\limits_{m,j=0}^{+\infty }\int\limits_{\mathbb{R}}\frac{%
P_{m}^{\left( \frac{1}{2}\gamma \right) }(x,\theta) }{\sqrt{%
\sigma _{\varepsilon }^{\beta ,\gamma }(m) }}\frac{P_{j}^{\left(
\frac{1}{2}\gamma \right) }(x,\theta) }{\sqrt{\sigma
_{\varepsilon }^{\beta ,\gamma }(j) }}\mid \psi _{m}^{\gamma
,\beta }><\psi _{j}^{\gamma ,\beta }\mid \left( \mathcal{N}_{\theta ,\gamma
,\beta ,\varepsilon }(x) \right) ^{-1}d\mu _{\theta ,\gamma
,\beta ,\varepsilon }(x) \right) [\varphi]
\label{4.30}
\end{align}
Replace $d\mu _{\theta ,\gamma ,\beta ,\varepsilon }(x) =%
\mathcal{N}_{\theta ,\gamma ,\beta ,\varepsilon }(x) \Upsilon
_{\theta ,\gamma }(x) dx,$ then Eq. (4.30) takes
the form
\begin{equation}
\mathcal{O}_{\theta ,\gamma ,\beta ,\varepsilon
}=\sum\limits_{m,j=0}^{+\infty }\left[ \int\limits_{\mathbb{R}}\frac{%
P_{m}^{\left( \frac{1}{2}\gamma \right) }(x,\theta) }{\sqrt{%
\sigma _{\varepsilon }^{\beta ,\gamma }(m) }}\frac{P_{j}^{\left(
\frac{1}{2}\gamma \right) }(x,\theta) }{\sqrt{\sigma
_{\varepsilon }^{\beta ,\gamma }(j) }}\Upsilon _{\theta ,\gamma
}(x) dx\right] \mid \psi _{m}^{\gamma ,\beta }><\psi
_{j}^{\gamma ,\beta }\mid   \label{4.31}
\end{equation}
Then, we need to consider the integral
\begin{equation}
I_{m,j}\left( \theta ,\gamma ,\varepsilon \right) :=\int\limits_{\mathbb{R}%
}P_{m}^{\left( \frac{1}{2}\gamma \right) }(x,\theta)
P_{j}^{\left( \frac{1}{2}\gamma \right) }(x,\theta) \Upsilon
_{\theta ,\gamma }(x) dx  \label{4.32}
\end{equation}
We recall the orthogonality relations of the Meixner-Pollaczek polynomials (\cite[p. 764]{14}):
\begin{equation}
\int\limits_{\mathbb{R}}P_{m}^{(\nu) }(x,\theta)
P_{j}^{(\nu) }(x,\theta) \omega _{\nu }\left(
x,\theta \right) dx=\frac{\Gamma (2\nu +m) \cos ec\theta }{m!}%
\delta _{m,j}  \label{4.33}
\end{equation}
where
\begin{equation}
\omega _{\nu }(x,\theta) =\frac{(2\sin \theta)
^{2\nu -1}}{\pi \cos ec(\theta )}e^{-( \pi -2\theta)
x}\left| \Gamma (\nu +ix) \right| ^{2}.  \label{4.34}
\end{equation}
This suggests us to set
\begin{equation}
\Upsilon _{\theta ,\gamma }(x) :=\frac{\omega _{\frac{1}{2}%
\gamma }(x,\theta) }{\Gamma (\gamma) \cos ec(\theta )}.  \label{4.35}
\end{equation}
Therefore, (4.32) reduces to
\begin{equation}
I_{m,j}\left( \theta ,\gamma ,\varepsilon \right) =\frac{\left( \gamma
\right) _{m}}{m!}\delta _{m,j}  \label{4.36}
\end{equation}
which means that the operator in (4.31) takes the form:
\begin{align}
\mathcal{O}_{\theta ,\gamma ,\beta ,\varepsilon }\equiv \mathcal{O}_{\gamma
,\beta ,\varepsilon } &=\sum\limits_{m,j=0}^{+\infty }\frac{1}{\sqrt{\sigma
_{\varepsilon }^{\beta ,\gamma }(m) }\sqrt{\sigma _{\varepsilon
}^{\beta ,\gamma }(j) }}\frac{(\gamma) _{m}}{m!}%
\delta _{m,j}\mid \psi _{m}^{\gamma ,\beta }><\psi _{j}^{\gamma ,\beta }\mid
\label{4.37}
\\
&=\sum\limits_{m=0}^{+\infty }\frac{(\gamma) _{m}}{m!}\frac{1}{%
\sigma _{\varepsilon }^{\beta ,\gamma }(m) }\mid \psi
_{m}^{\gamma ,\beta }><\psi _{m}^{\gamma ,\beta }\mid   \label{4.38}
\end{align}
Recalling the expression
\begin{equation}
\sigma _{\varepsilon }^{\beta ,\gamma }(m) :=\left( m!\right)
^{-1}(\gamma) _{m}e^{2\beta (2m+\gamma)
\varepsilon },  \label{4.39}
\end{equation}
we arrive at \qquad \
\begin{equation}
\mathcal{O}_{\theta ,\gamma ,\varepsilon }[\varphi]
=\sum\limits_{m=0}^{+\infty }e^{-2\beta (2m+\gamma) \varepsilon
}\left( \mid \psi _{m}^{\gamma ,\beta }><\psi _{m}^{\gamma ,\beta }\mid
\right) [\varphi] .  \label{4.40}
\end{equation}
For $u\in \mathbb{R}_{+},$ we can write
\begin{align}
\mathcal{O}_{\theta ,\gamma ,\varepsilon }[\varphi] (u) &=\sum\limits_{m=0}^{+\infty }e^{-2\beta (2m+\gamma)
\varepsilon }<\varphi \mid \psi _{m}^{\gamma ,\beta }>\psi _{m}^{\gamma
,\beta }(u)   \label{4.41}
\\&
=\sum\limits_{m=0}^{+\infty }e^{-2{\beta }(2m+\gamma)
\varepsilon }\left( \int\limits_{0}^{+\infty }\varphi (\xi)
\psi _{m}^{\gamma ,\beta }(\xi) d\xi \right) \psi _{m}^{\gamma
,\beta }(u)   \label{4.42}
\\&
=\int\limits_{0}^{+\infty }\varphi (\xi) \left(
\sum\limits_{m=0}^{+\infty }e^{-2\beta (2m+\gamma) \varepsilon
}\psi _{m}^{\gamma ,\beta }(\xi) \psi _{m}^{\gamma ,\beta
}(u) \right) d\xi   \label{4.43}
\end{align}
We are then lead to calculate the sum
\begin{equation}
\mathcal{G}_{\varepsilon }^{\alpha ,\beta }\left( u,\xi \right)
:=\sum\limits_{m=0}^{+\infty }e^{-2\beta (2m+\gamma)
\varepsilon }\psi _{m}^{\gamma ,\beta }(\xi) \psi _{m}^{\gamma
,\beta }(u) .  \label{4.44}
\end{equation}
For this we recall the explicite expression of the wavefunction $\psi
_{m}^{\gamma ,\beta }(\xi) $ in $\left( 2.9\right) $. So that
the above sum reads
\begin{align}
\mathcal{G}_{\varepsilon }^{\alpha ,\beta }\left( u,\xi \right) &=e^{-2\beta
\gamma \varepsilon }2\beta ^{\gamma }\left( \xi u\right) ^{\gamma -\frac{1}{2%
}}e^{-\frac{1}{2}\beta \left( u^{2}+\xi ^{2}\right) }  \nonumber
\\&
\times \sum\limits_{m=0}^{+\infty }\mu ^{m}\frac{m!}{\Gamma \left( m+\gamma
\right) }L_{m}^{\left( \gamma -1\right) }\left( \beta u^{2}\right)
L_{m}^{\left( \gamma -1\right) }\left( \beta \xi ^{2}\right) .  \label{4.45}
\end{align}
Eq. (4.45) can be rewritten as
\begin{equation}
\mathcal{G}_{\varepsilon }^{\alpha ,\beta }\left( u,\xi \right) =e^{-2\beta
\gamma \varepsilon }2\beta ^{\gamma }\left( u\xi \right) ^{\gamma -\frac{1}{2%
}}e^{-\frac{1}{2}\beta \left( u^{2}+\xi ^{2}\right) }K\left( e^{-4\beta
\varepsilon };u,\xi \right)   \label{4.46}
\end{equation}
where we have introduced the kernel function
\begin{equation}
K\left( \rho ;u,\xi \right) :=\sum\limits_{m=0}^{+\infty }\rho ^{m}\frac{m!}{%
\Gamma \left( m+\gamma \right) }L_{m}^{\left( \gamma -1\right) }\left( \beta
u^{2}\right) L_{m}^{\left( \gamma -1\right) }\left( \beta \xi ^{2}\right)
,0<\rho <1.  \label{4.47}
\end{equation}
Returning back to Eq. (4.43) and taking into account Eq. (4.47, we get that
\begin{equation}
\mathcal{O}_{\theta ,\gamma ,\varepsilon }[\varphi] \left(
u\right) =2\beta ^{\gamma }e^{-2\beta \varepsilon }u^{\gamma -1}e^{-\frac{1}{%
2}\beta u^{2}}\int\limits_{0}^{+\infty }\xi ^{\gamma -\frac{1}{2}}e^{-\frac{1%
}{2}\beta \xi ^{2}}K\left( e^{-4\beta t},\beta u^{2},\beta \xi ^{2}\right)
\varphi (\xi) d\xi   \label{4.48}
\end{equation}
We split the right hand side of Eq. (4.48)
\begin{equation}
\mathcal{O}_{\theta ,\gamma ,\varepsilon }[\varphi] \left(
u\right) =\vartheta _{\beta ,\gamma ,\varepsilon }(u) M\left[
\varphi \right] (u) ,  \label{4.49}
\end{equation}
where
\begin{equation}
M[\varphi] (u) =\frac{1}{2}\beta ^{-\frac{1}{2}%
\gamma -\frac{1}{4}}\int\limits_{0}^{+\infty }K\left( \rho ;w,s\right)
h\left( s\right) s^{\gamma -1}e^{-s}ds,  \label{4.50}
\end{equation}
with $w:=\beta u^{2}$, and
\begin{equation}
h\left( s\right) :=s^{-\frac{1}{2}\gamma +\frac{1}{4}}e^{\frac{1}{2}%
s}\varphi \left( \beta ^{-\frac{1}{2}}\sqrt{s}\right) .  \label{4.51}
\end{equation}
By direct calculations, one can check that $h\in L^{2}\left( \mathbb{R}%
_{+},s^{\gamma -1}e^{-s}ds\right) .$ Precisely, we have that
\begin{equation}
\left\| h\right\| _{L^{2}\left( \mathbb{\,R}_{+},s^{\gamma
-1}e^{-s}ds\right) }^{2}=2\sqrt{\beta }\left\| \varphi \right\|
_{L^{2}\left( \mathbb{R}_{+}\right) }  \label{4.52}
\end{equation}
We are now in position to apply the result of B. Muckenhoopt \cite{15} who considered the Poisson integral of a function $f\in
L^{p}\left( \mathbb{R}^{+},s^{\eta }e^{-s}ds\right) ,\eta >-1,1\leq p\leq
+\infty $ defined by
\begin{equation}
A\left[ f\right] \left( \rho ,w\right) :=\int\limits_{0}^{+\infty }K\left(
\rho ,w,s\right) f\left( s\right) s^{\eta }e^{-s}ds,0<\rho <1  \label{4.53}
\end{equation}
with the kernel $K\left( \rho ,w,s\right) $ defined as in $\left(
4.47\right) .$ He proved that $\lim_{\rho \rightarrow 1^{-}}A\left[ f\right]
\left( \rho ,x\right) =f(x) $ almost everywhere in $[0,+\infty [ ,1\leq p\leq \infty .$ We apply this result in the case $%
p=2,f=h$ and $A\equiv M$ to obtain that
\begin{equation}
M[\varphi] (u) \rightarrow \frac{1}{2}\beta ^{-%
\frac{1}{2}\gamma -\frac{1}{4}}h( \beta u^{2}) =\beta ^{-\frac{1}{%
2}\gamma +\frac{1}{4}}u^{-\gamma +\frac{1}{2}}e^{\frac{1}{2}\beta
u^{2}}\varphi (u)   \label{4.54}
\end{equation}
Recalling that $\rho =e^{-4\beta t},$ we get that
\begin{equation}
\mathcal{O}_{\theta ,\gamma ,\varepsilon }[\varphi] \left(
u\right) =\vartheta _{\beta ,\gamma ,\varepsilon }(u) M\left[
\varphi \right] (u) \rightarrow \varphi (u) \text{
as }\varepsilon \rightarrow 0^{+}  \label{4.55}
\end{equation}
which means that
\begin{equation}
\lim_{\varepsilon \rightarrow 0^{+}}\int\limits_{\mathbb{R}}\mid
x,\varepsilon ><x,\varepsilon \mid d\mu _{\theta ,\gamma ,\varepsilon
}(x) =\mathbf{1}_{L^{2}(\mathbb{R}_{+},d\xi) .}
\label{4.56}
\end{equation}
This ends the proof.
\fin \\

\section{Summary}

We have been a concerned with the model of the Gol'dman-Krivchenkov
Hamiltonian acting on the Hilbert space of square integrable functions on
the positive real half-line. We have constructed a family of coherent states
labeled by points of the whole real line, depending on some parameters and
belonging to the state Hilbert space of this Hamiltonian. We have adopted a
generalized formalism of canonical coherent states when written as
superpositions of the harmonic oscillator number states. That is, we have
presented our generalized coherent states as a superposition of an
orthonormal basis of the state Hilbert space of the Hamiltonian. The
Meixner-Pollaczek polynomials have been chosen to play the role of
coefficients in this superposition. This choice enables us to present the
constructed states in a closed form. The state Hilbert space identity is
solved by an new way as a limit with respect to a certain parameter by
exploiting the results of  B. Muckenhoupt on Poisson integrals for Laguerre
expansions. The obtained result will lead to investigate the class of
special functions and orthogonal polynomials that could play the role of
\textit{coefficients} in the series expansion of coherent states together
with the class of functions that could be considered as \textit{number states%
} for a certain Hamiltonian in a such way that the present procedure works.


\section*{References}
\begin{thebibliography}{15}
\bibitem{1}  V. V. Dodonov, 'Noncalssical' states in quantum optics: a
'squeezed review of the first 75 years, \textit{J.opt.B:Quantum
Semiclass.opt. }\textbf{4}, R1-R33 (2002)

\bibitem{2}  S. T. Ali, J.P. Antoine and J.P. Gazeau, Coherent states,
Wavelets and their Generalizations, New York Springer, 2000

\bibitem{3}  J. R. Klauder and B. S. Skagerstam, Coherent states,
Applications in Physics and Mathematical physics, Singapore, World
Scientific, 1985

\bibitem{4}  A. M. Perelomov, Generalized coherent states and their
Applications, Berlin, Springer, 1986

\bibitem{5}  J. P. Gazeau, Coherent States in Quantum Physics,Wiley Verlag,
2009

\bibitem{6}  I. I. Gol'dman and D. V.Krivchenkov, Problems in Quantum
Mechanics, Pergamon, London, 1961

\bibitem{7}  R. L. Hall, N. Saad and A.B. Von Keviczky, Closed-form sums for
some perturbation series involving associated Laguerre polynomials, \textit{%
J. Phys. A: Math. Gen. }\textbf{34 }(2001), 11287-11300

\bibitem{8}  W.Magnus, F.Oberhettinger \& R.P.Soni, Formulas and Theorems
for the Special Functions of Mathematical Physics, Springer-Verlag Berlin
Heidelberg New York, 1966.

\bibitem{9}  R.L. Hall, N. Saad \& A.B. von Kevicsky, Spiked harmonic
oscillators,\textit{\ arXiv: math-ph/ 0109014v1, 18 sep 2001}

\bibitem{10}  S. M. Nagiev, E. I. Jafarov and R. M. Imanov, On a dynamical
symmetry of the relativistic linear singular oscillator, \textit{arXiv:
math-ph/0608057v2, }28 sep 2006

\bibitem{11}  K. Thirulogasantar and N. Saad: Coherent states associated
with the wavefunctions and the spectrum of the isotonic oscillator, \textit{%
J. Phys. A: Math. Gen. }\textbf{37 }(2004), 4567-4577.

\bibitem{12}  M.E.H. Ismail and D. Stanton, Classical orthogonal polynomials
as moments, \textit{Can. J. Math}.\textbf{\ 49} (3) pp.520-542, 1997

\bibitem{13}  E. D. Rainville, Special functions, The Macmillan company, New
York, 1963 page 213

\bibitem{14}  P. A. Lee, Probabilistic derivation of a bilinear summation
formula for the Meixner-Pollaczek polynomials,\textit{\ Internat. J. Math.
\& Math. Sc}i. Vol. 3 No. 4 (1980) 761-771

\bibitem{15}  B. Muckenhoupt, Poisson integrals for Hermite and Laguerre
expansions, \textit{Trans. Amer. Math. So}c, \textbf{139,}pp. 231-242\textbf{%
,} 1969
\end{thebibliography}
\end{document}